\documentstyle[12pt,epsf]{article}

\input epsf
\if@twoside
 \else
 \fi

 \parskip \medskipamount
\newcommand{\be}{\begin{eqnarray}}
\newcommand{\ee}{\end{eqnarray}}

\begin{document}

\hbox{} \nopagebreak
\vspace{-3cm} \addtolength{\baselineskip}{.8mm} \baselineskip=24pt
\begin{flushright}
 CERN-TH/2001-063
\end{flushright}

\vspace{2.5cm}
\begin{center}
{\Large \bf  $Z_N$ wall junctions: Monopole fossils in hot QCD} \\
\vskip 0.3 cm 
 Alex Kovner\footnote{Alexander.Kovner@cern.ch}\\
\vspace{0.5 cm}
{\it Theory Division, CERN, CH-1211, Geneva 23, Switzerland}\\
\vskip 0.2 cm
and
\vskip 0.2 cm
{\it Department of Mathematics and Statistics, 
University of Plymouth,\\
Plymouth PL4 8AA, UK}\\

\end{center}  

\vspace*{1.5cm}


\begin{abstract}
\baselineskip=18pt  
We point out that 
the effective action of  hot Yang--Mills theories has
semi-classical solutions, which are naturally identified with monopole 
world lines, ``frozen'' into the short imaginary time dimension. The 
solutions look like wall junctions: 
lines along which $N$ electric $Z_N$ domain walls come
together. They are instrumental in reconciling explicit perturbative
calculations at high temperature with the magnetic $Z_N$ symmetry.
\end{abstract}
\vskip 3cm
CERN-TH/2001-063\\
February 2001

\vfill

\newpage

The idea that confinement in non abelian gauge theories is brought about
by condensation of magnetic monopoles is an old one \cite{duals}.
The monopole condensation has been shown to
cause confinement in abelian - like models, such as the (2+1)- dimensional
Georgi--Glashow model \cite{polyakov}, or perturbed $N=2$ supersymmetric 
gauge theory in 3+1 dimensions \cite{sw}.
However in a genuine non abelian context the monopoles have been rather elusive. 
Their identification through various abelian projections suffers from 
lack of gauge invariance and thus gauge dependence of various monopole 
properties \cite{abelianp}.

Things would be clear if the monopoles were to appear as classical 
solutions of the QCD equations of motion. This, however,
does not happen. This is not surprising by itself, since even if the 
monopoles do exist, they are strongly coupled and their density in the vacuum
is not expected to be low. Thus it is hard to expect that they appear as well 
formed classical solutions.

The situation is somewhat different at high temperature. There QCD 
is perturbative, or at least has a perturbative sector which is usually 
described in terms of the weakly coupled 
effective theory for the Polyakov loop. This effective theory, being 
weakly coupled, may sustain meaningful classical solutions.
The purpose of this note is to point out that such classical solutions exist, 
and that in a certain sense (to be explained below) they represent the 
``world lines'' of the dynamical QCD monopoles. These word-lines are space-like
and are squeezed in the short imaginary time direction.

Consider the high - temperature effective action of pure Yang--Mills theory in 
3+1 dimensions. It is written in terms of $N-1$ independent phase fields 
$A_\alpha$ ---
the phases of the eigenvalues of the Polyakov loop
(see for example \cite{weiss}). To one loop
this effective action is
\be
S={1\over 2g^2N}\sum_{ij}(\partial_\mu A_{\alpha\beta})^2+{2\over 3}
T^4\pi^2\sum_{ij}B_4(A_{\alpha\beta}).
\label{action}
\ee
Here $A_{\alpha\beta}={A_\alpha-A_\beta\over 2\pi T}$, with 
$\alpha,\beta=1,..., N$, and $A_{N}=-\sum_{\alpha=1}^{N-1}A_\alpha$. The
potential $B_4$ is the Bernoulli polynomial.
Since $A_i$ are phases, the first homotopy group of the field manifold is
non-trivial, and thus there must be stable classical configurations with 
nontrivial winding. 
Let us require that all the phases $A_\alpha$, 
$\alpha=1,...,N-1$, wind once when going around
some straight line $C$. If we disregard the potential term in 
eq. (\ref{action}), the equations of motion obviously have a solution with 
this boundary condition
\be
A_\alpha(x)=\theta(x_{perp}),
\ee
where $x_{\perp}$ are the coordinates perpendicular to $C$.
The action per unit length of this solution is logarithmically divergent.
Reintroducing the potential in the action, we see that 
the solution changes significantly. 
Now that the fields $A_\alpha$ are massive it is energetically favourable to
concentrate the winding within a two-dimensional surface of finite width, 
rather than have it delocalized in the whole space. The solution therefore
will have a wall-like structure, where the fields $A_\alpha$ 
vary within a width
$M_D^{-1}$ of half a plane with boundary $C$. In fact the finer 
structure of the solution can be understood quite easily. The potential
in eq. (\ref{action}) has $N$ minima, 
$A_\alpha=2\pi n/N$, for $n=0, 1, ..., N-1$.
While varying within the wall, the fields $A_\alpha$ 
have to pass through all these
minima. Since the interaction of $A_\alpha$ is weak (especially at large $N$), 
the wall will split into $N-1$ vacuum regions separated by the
standard $Z_N$ domain walls \cite{znwalls}\footnote{Of course if the boundary
$C$ is one straight line, the walls will repel each other and will spread 
like a fan. We have in mind however that $C$ is closed, albeit its curvature 
is very small on the scale of $M_D^{-1}$.}
. This {\it N}-layered 
``sandwich'' solution 
is depicted in Fig. 1. One can think of it as a wall junction, where $N$ domain
walls come together at the line $C$.

\begin{figure}
\begin{center}
\epsfxsize=5in
\epsfbox{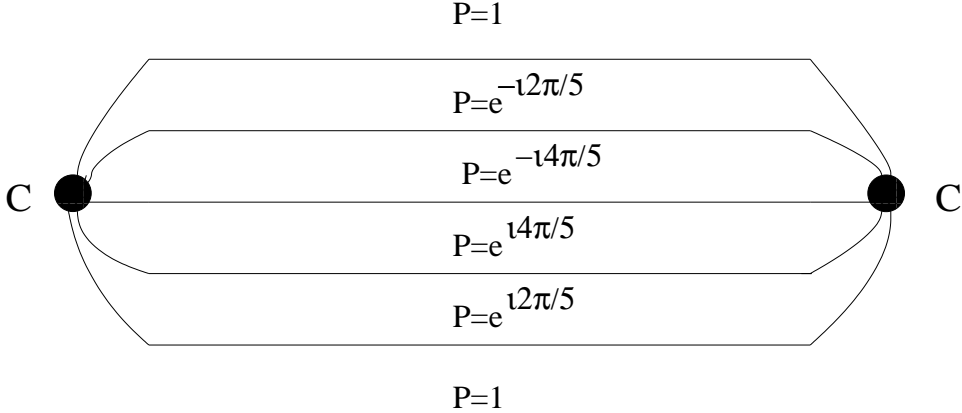}
\end{center}
\caption{The classical monopole fossil  configuration ---
the junction of $N$ $Z_N$ domain walls (here $N=5$). The values of 
the trace of the Polyakov loop in each of the vacua
in the ``sandwich'' are indicated. The contour $C$
is perpendicular to the plane of the figure.}
\end{figure}

The action per unit length is now linearly divergent
\be
S_{mf}=N\tilde\sigma L\ \ \ ,
\ee
where $\tilde\sigma$ is the $Z_N$ domain wall tension.
\be
\tilde\sigma={4\pi^2T^2(N-1)\over 3{\sqrt{3g^2(T)N}}}\ \ \ .
\ee

If instead of the straight line we choose a closed curve for $C$, the
sandwich will be finite and will span the minimal area subtended by $C$.

What is the physical meaning of this solution? 
Imagine for a second that at zero temperature there are indeed monopoles. 
The charge of such a monopole would be $N$ units of the fundamental 
magnetic charge allowed by the Dirac quantization conditions. Consider a
word line of such a monopole with the unit tangential vector parallel to
some coordinate axis $x_\lambda$.
 In the directions perpendicular to the world line,
the monopole configuration has the Coulomb magnetic field  
$\tilde F_{\lambda\nu}={N\over g}{x_\nu\over x^3}$, $\nu\ne \lambda$.
The action of such a configuration is proportional to the length of the 
word line, with the proportionality coefficient equal to the Coulomb energy 
of the monopole. Now let us increase the temperature. In the imaginary time 
formalism this amounts to making one direction finite
with the size $\beta$. Let us take this 
finite direction to be perpendicular to $x_\lambda$. Because of the periodic
boundary conditions in this finite direction, the magnetic flux 
cannot penetrate the boundary. The magnetic field lines will therefore 
bend as they come close to the boundary. Thus 
at distances
larger than $\beta$ the whole magnetic flux will
be effectively squeezed into two transverse directions.
At these distances the field will be two-dimensional Coulomb, 
rather than three-dimensional 
Coulomb, and the action density per unit length will
logarithmically diverge with the volume of the system. This is entirely
analogous to the change of the profile and the interactions of the
instantons in (2+1)-dimensional Georgi--Glashow model \cite{gg, ggi}.
Also, since the compact dimension is the imaginary time, the components of 
the dual field strength which do not vanish in this configurations are the
ones perpendicular to the time axis. Thus those are really electric fields
rather than magnetic fields, and are representable in terms of the
scalar potential $A_0$. Remembering that our fields $A_\alpha$ 
are indeed the scalar potentials, we expect these configurations to appear
directly in our effective action. Indeed, those are precisely the 
classical solutions described above. Indeed, disregarding the 
Debye mass term in the effective Lagrangian, the action per unit length of
our solution is precisely the two-dimensional Coulomb energy of a monopole
with magnitude ${N\over g}$. 
The presence of the Debye mass above the deconfining transition is the 
reflection of the restoration of the magnetic $Z_N$ symmetry \cite{zn}, and 
it naturally affects the action of classical solutions.
Again the situation is extremely similar to the case of (2+1) dimensions \cite{gg, ggi}.
The squeezed instantons interact logarithmically at low (but non-zero)
temperature, but the interaction becomes linear above the deconfining 
transition.

Our classical solutions are therefore just the (space-like) worldliness 
of magnetic monopoles squeezed and preserved in the compact imaginary
time dimension. Hence we will refer to these solutions as monopole fossils.

Since the world lines are space-like, these objects strictly speaking
do not represent 
physical monopoles, but rather are related to 
magnetic vortices. To see this explicitly, let us consider the calculation
of the expectation value of 
the $N$-fold 't Hooft loop \cite{thooft}, -
the operator that creates a magnetic vortex with flux $2\pi N/g$:
\begin{equation}
V_N(C)=\exp\big\{{2\pi i\over g}\int_S d^2 S^i {\rm Tr}YE^i\big\}
\label{v}
\end{equation}
where the hypercharge generator $Y$ is defined as
\begin{equation}
Y={\rm diag} \left(1,1,...,-(N-1)\right)
\end{equation}
and the integration goes over a surface $S$ bounded by the curve $C$. 
The operator does not depend on which surface $S$ is chosen as long as its 
boundary is fixed \cite{zn,kovner}. 
This operator is the operator of a singular gauge transformation
in the hypercharge direction. The nature of the 
singularity is such that the gauge phase winds once when 
encircling
the contour $C$.
The path integral representation for this expectation value is then
precisely the same as for the vacuum partition function except for the
``boundary condition'' imposed on $A_\alpha$ in the integration domain:
it must have a unit 
winding relative to $C$. Thus the steepest descent
calculation of this expectation value is dominated by the classical solution
we have just discussed.

The monopole fossils are instanton-like objects, which describe the
process of creation of a magnetic vortex with magnetic flux ${2\pi\over g}$.
To avoid confusion we wish to make the following comment. The operator
$V_N$ defined in eq. (\ref{v}) is in fact a trivial operator. 
Non-perturbatively,
it is equivalent to the unit operator.
This is simplest to understand by noting that it commutes with all gauge- 
invariant operators \cite{zn}.
The reason this is not obvious in our derivation is that we implicitly 
assumed that the integration in the thermal 
path integral is perturbative and is thus only over the small fields
$A_\alpha$. 

This is not to say that the monopole fossils are not important. Their
importance is precisely in restoring the triviality of the $N$-fold
vortex operator within the perturbative/semiclassical domain. In this they are
directly analogous to usual instantons in QCD. Recall that if we neglect
instantons in the QCD vacuum path integral, the operator of a large gauge 
transformation  $U$ is non-trivial. 
For example its insertion into the path integral
changes the boundary conditions on the gauge fields by changing the
total topological charge by one unit. Thus
perturbatively
the calculation of $\langle U\rangle$ 
is dominated by the one instanton configuration
and the result would be $\exp \{-S_{inst}\}$. It is only after we sum over all 
``dynamical'' 
instantons and anti-instantons in the path integral that the triviality of $U$
is restored. Technically this is because the vev of $U$ will
be dominated by a configuration where a ``dynamical'' anti-instanton from the
vacuum ensemble will sit on top of the instanton induced by the explicit
insertion of $U$. Thus the leading contribution will be $1$ rather than
 $\exp \{-S_{inst}\}$. 
More generally, in calculation of any correlation function
insertions of $U$ are unimportant. Such an insertion amounts to adding one more
instanton to the ensemble which, as it is, has an indeterminate number 
of randomly distributed instantons and anti-instantons. 
Thus the one extra insertion does not change the ensemble, and this restores 
the triviality of the operator $U$ in such a semi-classical context.

Exactly the same thing happens with our monopole fossils. The perturbative 
result 
$\langle V_N(C)\rangle=\exp\{-S_{mf}\}$ gets changed to $\langle V_N\rangle=1$ once we sum
over all possible monopole fossils in the thermal ensemble, since there 
is always a ``dynamical'' fossil to screen the one induced by the insertion 
of $V$. Note that the action of a large monopole fossil is very large
(proportional to the area with large tension),
and thus the fossils are not abundant in the vacuum ensemble. This does 
not prevent them from dominating the path integral for $\langle V_N\rangle$, 
just as
the instantons with large action dominate $\langle U\rangle$.

Another quantity in which the contributions of fossils are important
are expectation values of $k$-fold 't Hooft loops with $k<N$. The calculation
of these expectation values has been performed recently in \cite{giovan}
with the result
\be
\langle V_k(C)\rangle
=\exp\{-\tilde\sigma_kS\}
\label{k}
\ee
with
\be
\tilde\sigma_k={N-k\over N-1}k\tilde\sigma\ \ \ .
\label{sigmak}
\ee
The main feature of this result which is of interest to us, is that 
$\langle V_k(C)\rangle=\langle V_{N-k}(C)\rangle$. 
This is of course what one expects non-perturbatively.
For large areas, $\langle V_k(C)\rangle$ 
measures the longest correlation length in the channel
with magnetic flux $2\pi k/g$. Since the global magnetic symmetry in the theory
is $Z_N$, it must be true that the same correlation length also appears in
the channel with flux $2\pi (k-N)/g$. 
On the other hand if the symmetry were $U(1)$ no such relation
would hold.
Thus eq. (\ref{sigmak}) is consistent with the magnetic symmetry 
being $Z_N$ rather than $U(1)$.

However, from the point of view of a purely perturbative calculation,
this result seems surprising. The interaction between two $k=1$ walls
is only via the phase fields $A_\alpha$. 
Such an interaction is usually repulsive.
In fact one can readily convince oneself that at least when the separation 
between the walls is larger than the inverse Debye mass, the
interaction is indeed repulsive. Thus, if we were to look for a stable
configuration with $k=2$ by just solving the classical equations with 
the boundary conditions corresponding to $k=2$, we would find that the 
stable solution is two walls fairly well separated in the transverse direction
with the action $2\tilde\sigma$. And the same would happen for any $k$ - 
the
classical action will be just proportional to $k$. The authors of \cite{giovan}
note explicitly that if they use the simple definition of the 
$k$-fold 't Hooft loop and perform straightforward 
perturbative calculation they indeed get this result.
\footnote{The 
actual calculation in \cite{giovan} is performed differently, by
globally sampling the space of allowed configurations of $A_\alpha$ and thus
avoiding this problem.}

This puzzle is resolved by the presence of the monopole fossil solutions.
Without their contributions, the perturbation theory
distinguishes sharply between $V_k$ and $V_{N-k}$, and thus
leads to different values for $\tilde\sigma_k$ and $\tilde\sigma_{N-k}$ 
string tensions. On the other hand if we do take them into account
it is easy to see that when calculating $\langle V_{N-k}(C)\rangle$
 a single anti-fossil
sitting right on top of the contour $C$ will turn  
$\langle V_{N-k}(C)\rangle$ into
 $\langle V_{-k}(C)\rangle$ 
(which by charge conjugation is equal to  $\langle V_{k}(C)\rangle$).
The fossil does not have to sit right on top of $C$ to give a significant 
contribution. Its position and shape should be integrated over.
The mixing between  $\langle V_{N-k}(C)\rangle$ and  
$\langle V_{-k}(C)\rangle$ that the fossils 
bring about must also reduce the lowest mass in this channel, and thereby
bring the perturbative result for $\tilde\sigma_k=k\tilde\sigma$
in agreement with eq. (\ref{sigmak}). 
We do not attempt the explicit calculation
of this effect here, but note that in weakly interacting theories 
in 2+1 dimensions a similar calculation has been performed in detail in 
\cite{ggi}. It is indeed shown there that the high-temperature remnants of
monopole-instantons have the effect of increasing the longest correlation
length in a channel with a given magnetic flux through precisely the same 
mixing mechanism as discussed above.

Finally we note that other classical configurations with 
nontrivial winding can be considered. 
For example, one can require that $A_1$ winds
around some contour, but the other $A_\alpha$, $\alpha=2, ... , N-1$, do not.
This solution also has the meaning of a monopole fossil, but this time
of a monopole which carries the flux only within a particular $U(1)$
subgroup. These configurations may also be important in sem- classical 
calculations of the type described above. 

To conclude, we pointed out in this note that the effective action of 
high-temperature Yang--Mills theory has classical solutions that are naturally
identified with the space-like world lines of magnetic monopoles squeezed
in the imaginary time direction. The action of these solutions is
proportional to the minimal area subtended by the world-line $C$. 
Such a solution looks 
like a junction of $N$ $Z_N$ domain walls terminating on $C$.
Their 
importance in the semi-classical context is to break the magnetic symmetry
down to $Z_N$ from the apparent $U(1)$ by mixing the correlation functions
of $V_k(C)$ and $V_{k-N}(C)$.

\leftline{\bf Acknowledgements}
 I thank Chris Korthals - Altes for interesting discussions.

\end{document}